\documentclass[epj]{svjour}
\pdfoutput=1
\usepackage{amssymb, amsmath}   

\usepackage{graphicx}

\newcommand{\ecls}{{\vec e}_{\mbox{\tiny CLS}}}
\newcommand{\eclss}[1]{{\vec e}_{\mbox{\tiny CLS}, #1}}
\newcommand{\ep}{{\vec e}_{\perp}}

\begin{document}
\title{Localization of weakly disordered flat band states}

\author{Daniel Leykam$^{1,2}$, Joshua D. Bodyfelt$^{3}$, Anton S. Desyatnikov$^{4,2}$, and Sergej Flach$^{5,3}$}
\institute{$^{1}$Division of Physics and Applied Physics, School of Physical and Mathematical Sciences, Nanyang Technological University, Singapore 637371, Singapore\\
$^{2}$Nonlinear Physics Centre, Research School of Physics and Engineering, The Australian National University, Canberra ACT 2601, Australia\\
$^{3}$New Zealand Institute for Advanced Study, Centre for Theoretical Chemistry \& Physics, Massey University,  0745 Auckland, New Zealand\\
$^{4}$Physics Department, School of Science and Technology, Nazarbayev University, 53 Kabanbay Batyr Ave.,
Astana, Kazakhstan\\
$^{5}$Center for Theoretical Physics of Complex Systems, Institute for Basic Science, Daejeon, Korea}

\date{\today}

\abstract{
Certain tight binding lattices host macroscopically degenerate flat spectral bands. Their origin is rooted in local symmetries of the lattice, with destructive interference leading to the existence of compact localized eigenstates. We study the robustness of this localization to disorder in different classes of flat band lattices in one and two dimensions. Depending on the flat band class, the flat band states can either be robust, preserving their strong localization for weak disorder $W$, or they are destroyed and acquire large localization lengths $\xi$ that diverge with a variety of unconventional exponents $\nu$, $\xi \sim 1/W^{\nu}$. 
}

\authorrunning{D. Leykam, J. D. Bodyfelt, A. S. Desyatnikov, \& S. Flach}

\maketitle

\section{Introduction}

Wave propagation through periodic media reveals many fascinating behaviours arising from the interplay between diffraction and scattering. Generically the propagating modes are Bloch waves sensitive to the symmetries and topologies imprinted by the underlying periodic potential. Applying either a coupled mode or tight binding approximation allows one to model a few bands of the dispersion relation with a discrete set of effective lattice equations. The characteristic length and frequency scales can vary strongly. Examples include electrons in crystals or artificial quantum dot arrays~\cite{mesoscopic_review}, ultracold atoms in optical lattices~\cite{bloch2008}, microwaves in dielectric resonator networks~\cite{bellec13,casteels2016}, light propagation in waveguide networks~\cite{christodoulides2003}, and the dynamics of hybrid light-matter exciton-polariton condensates~\cite{carusotto_review,polariton_review}. The width of a Bloch band corresponds to a certain amount of available kinetic energy which determines the wave group velocity and sensitivity to perturbations.

A particularly interesting situation arises when some of the dispersion bands become strictly {\sl flat}. Then any relevant perturbation such as nonlinear effects or disorder can lift the band's macroscopic degeneracy and qualitatively change the nature of the eigenstates~\cite{huber2010,vicencio2013,johansson2015,goda2006,Nishino,chalker10,fb0,ge_arxiv}.  Such flat bands (FB) were originally studied as lattice models constructed using techniques from graph theory~\cite{mielke,tasaki,richter,derzhko_review,dias2015,nandy2015}. More recently they have been realized in experiments with photonic lattices~\cite{moti_lieb,compact_lieb_1,compact_lieb_2,mukherjee2015,induced_lieb,induced_kagome,sawtooth} and structured microcavities for exciton-polariton condensates~\cite{masumoto2012,flat_exciton_polariton}. In these cases the vanishing transverse group velocity of the flat band allows the lattice to support compact eigenstates that are perfectly localized to several lattice sites, with exactly vanishing amplitude on all other sites~\cite{bergman2008,richter}. The origin of such a compact localized state (CLS) is destructive interference that prevents diffraction and effectively decouples it from the rest of the lattice. This leads to distortion-free image transmission through photonic lattices~\cite{image_transmission} or the formation of many independent (incoherent) polariton condensates~\cite{flat_exciton_polariton}.

Any disorder with strength $W$ inevitably spoils this destructive interference, coupling the CLS to the lattice's Bloch waves and enabling diffraction or coherent condensation to a finite width determined the disorder-induced Anderson localization length $\xi$~\cite{anderson_light}. In this work we are concerned with what happens in the weak disorder limit as $W \rightarrow 0$. Does $\xi$ approach the size of a CLS, signifying their stability against weak disorder? Or does $\xi$ diverge, as one would expect for the Anderson localization length in a finite width Bloch band? Recent studies suggest that the behaviour strongly depends on the structure of the FB and its CLS; some CLS are robust, while others are highly sensitive to any disorder.

A recently proposed ``detangling'' procedure offers a systematic way to determine the response of CLS to disorder in any flat band lattice, by first transforming the model to a basis where in the absence of disorder the CLS appear as defects completely decoupled from the remaining dispersive Bloch bands~\cite{flach14}. In this basis, a generic disorder potential is split into two parts defined by the local symmetries of the CLS. The symmetric part maintains the compactness of the CLS while lifting their degeneracy, while the antisymmetric part hybridizes them with the continuum of dispersive degrees of freedom, analogous to a Fano resonance~\cite{miroshnichenko10}. Dispersive degrees of freedom near the FB energy are therefore scattered by a random collection of resonant scatterers, leading to a resonant enhancement of the effective disorder potential. The result for a particular 1D cross-stitch lattice was the localization length $\xi$ scaling with diagonal disorder strength $W$ as $\xi \sim 1/W$ at the FB energy -- diverging, but at variance with the familiar $\xi\sim 1/W^2$ law for Anderson localization in the other Bloch bands~\cite{flach14}.

In this work, we apply this detangling idea~\cite{flach14} to a variety of different quasi-1D and 2D tight binding lattices, and consider both diagonal and off-diagonal disorder, corresponding to random site energies and coupling strengths respectively. We find that in all cases the effective disorder felt by dispersive waves is resonantly enhanced by the CLS. Thereby, weak disorder potentials with bounded probability distribution functions become effective disorder that is strong and unbounded, with divergent variance and heavy tails resembling the Cauchy distribution. This divergent variance leads to a breakdown of standard weak disorder perturbation techniques and a violation of single parameter scaling~\cite{altshuler00,deych2003}. Computing the scaling of the flat band's Anderson localization length with weak disorder, we find the CLS are only stable (i.e. finite $\xi$ in limit $W \rightarrow 0$) when separated from other dispersive bands by a gap. Otherwise the localization length diverges with a whole variety of unconventional exponents sensitive to the class of the FB. We show that in 2D lattices the CLS-induced Fano resonances can no longer trap and localize waves, but resonances still affect the scaling of other measures of localization such as the eigenmode participation number. 

We start by reviewing the procedure for ``detangling'' flat band lattices into a basis formed by their compact localized states in Sec.~\ref{sec:review}. We show in Sec.~\ref{sec:disorder} that Anderson localization under the resulting Cauchy-tailed disorder can be understood through analogy with the exactly solvable Lloyd model, which predicts unconventional scaling of the localization length with the disorder strength. We then apply the detangling to four 1D example lattices in Sec.~\ref{sec:examples}, verifying against numerical computation of the localization length scaling. Sec.~\ref{2d} discusses the generalization to 2D systems. We summarize our results in Sec.~\ref{conclusion}.

\section{Flat band models, compact localized states, and detangling}
\label{sec:review}

We study a lattice wave eigenvalue problem of the type
\begin{equation} 
E \Psi_n = \epsilon_n \Psi_n -\sum_m t_{nm} \Psi_m, \label{eq:general_hamiltonian}
\end{equation}
where the wave components $\Psi_n$ are complex scalars allocated to points on a periodic lattice, the matrix $t_{nm}$ defines an interaction between them, and $\epsilon_n$ are onsite energies. The eigenvectors of such a tight binding model are delocalized Bloch waves with eigenenergies $E_{\nu}(\vec{k})$ forming a band structure, where $\vec{k}$ is the Bloch wavevector and $\nu=1,...,\mu$ is the band index. Excluding the trivial case of just one band, $\mu=1$, we consider models with at least one FB for which $E_{\nu}(\vec{k}) = const$. Due to the macroscopic degeneracy of the FB, its Bloch waves may be mixed to obtain localized FB eigenvectors (analogous to the construction of Wannier functions~\cite{wannier_review}). It is tempting to search for cases supporting \emph{compact} localized eigenvectors that are strictly zero outside a finite interval~\cite{richter}, although no general theorem guarantees their existence. When they do exist, such compact localized states can be further classified by the number of unit cells they span, $U$. 

\begin{figure}

\includegraphics[width=\columnwidth]{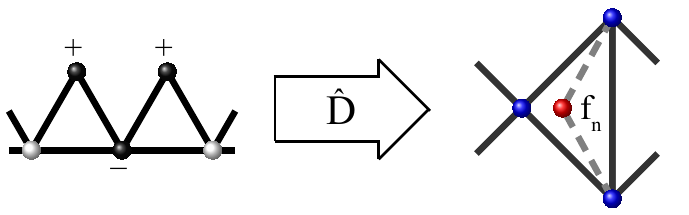}

\caption{(Color online) Schematics of the detangling procedure. Left: the 1D sawtooth lattice. A compact localized state of its flat band excites three sites with amplitudes as indicated by the $\pm$ signs. A local rotation $\hat{D}$ transforms the lattice to its partially diagonalized ``detangled'' version (right), consisting of decoupled compact localized states $f_n$ that only become coupled to the lattice via perturbations (dashed links), whereas the black links indicate hopping terms (not necessarily uniform) that are nonzero in the absence of perturbations.}

\label{fig:detangling}

\end{figure}

The nonzero amplitudes $\Psi_n$ of each CLS define a direction in the vector space formed by the lattice wave amplitudes. A rotation $\hat{D}$ can thus be performed to produce a new basis vector $\ecls$ aligning with the CLS direction, along with a remaining set of orthogonal basis vectors $\ep$. This local rotation, which respects the symmetry of a given CLS, literally rotates the CLS degree of freedom out of the lattice, as illustrated in Fig.~\ref{fig:detangling}. When starting with a network of $M$ lattice points and $N < M$ unit cells, applying this transformation will produce one ``detangled'' CLS satisfying a trivial equation of the type $E \Psi_\mathcal{S} = E_{FB} \Psi_\mathcal{S}$ and $M-1$ equations for the remaining network. In effect, the Hamiltonian is partially diagonalized; the CLS is decoupled from all other sites in this new basis.

The benefit of applying such a transformation is that perturbations to the original Hamiltonian network can then be split into a symmetric part which maintains the existence of the CLS, but possibly shifting its energy, and an antisymmetric part that couples it into the remaining degrees of freedom (dispersive band states, and possibly a finite number of neighbouring CLS). In particular, we immediately observe that this antisymmetric coupling is \emph{local} in space due to the compact nature of the CLS. 

For $U=1$ the set of all CLS (note that there is always one CLS per flat band and unit cell) forms an orthonormal complete basis for the flat band. One can always find a detangling transformation that decouples all the CLS simultaneously, implying the corresponding flat band projector operator is highly reducible in position space~\cite{mielke2012}. 

When $U>1$ the CLS lack orthogonality, but they may still form a complete basis for the flat band. If the CLS are incomplete, the missing states required to complete the flat band basis can be accounted for e.g. by using graph theoretical arguments and Euler's theorem~\cite{mielke,bergman2008}. One can perform a similar decoupling by considering a superlattice formed by groups of $U$ unit cells. Using such an enlarged unit cell, $N/U$ of the flat band states become CLS of class $U=1$. However there is no basis that can simultaneously decouple the remaining flat band states; the flat band projector operator is irreducible in position space. Nevertheless in many cases this partial detangling of the flat band is already sufficient to understand the impact of perturbations such as disorder, especially when the perturbations destroy the translational invariance of the lattice.

\section{Disorder}
\label{sec:disorder}

Now consider the effect of disorder on the above lattices, such that onsite potentials are uncorrelated, uniformly distributed random variables, $\epsilon_n \in [-W/2,W/2]$. Generically, the symmetric part of the disorder will smear out the energy of the CLS to a width $W$ while retaining their compactness. The antisymmetric part couples the CLS to the other dispersive states, which are typically Anderson-localized to a size of $\xi \sim 1/W^2$~\cite{kramer93} in 1D systems, and potentially to neighbouring CLS and to CLL.

To understand the effect of coupling the disordered CLS to the dispersive states, we propose a simple model generalizing the Fano-Anderson model of Ref.~\cite{miroshnichenko10}. For sufficiently weak disorder $W \ll t$, where $t$ is the scale of the hopping terms in Eq.~\ref{eq:general_hamiltonian} (which we normalize to 1) first order perturbation theory
 applies. The energy $E_j$ of the $j$th eigenvector $\Psi_n^{(j)}$ is shifted by $\delta E_j = \sum_n \Psi_n^{(j)*} \epsilon_n \Psi_n^{(j)} \sim W$, while the eigenvectors are perturbed by
\begin{equation}
\delta \Psi_n^{(j)} = \sum_{j^{\prime}\ne j} \frac{\sum_m \Psi_m^{(j^{\prime})*} \epsilon_m \Psi_m^{(j)}}{E_j - E_{j^{\prime}}} \Psi_n^{(j^{\prime})}.
\end{equation}
Since $W \ll 1$ these corrections are negligibly small, except for near-degenerate (i.e. resonant) states with $|E_j - E_{j^{\prime}}| < W$, which is only satisfied by the degenerate flat band states and possibly some states belonging to the nearest dispersive band. Therefore a two band approximation to this near-degenerate subspace is justified. Within this subspace we approximate the eigenvalue problem as
\begin{eqnarray}
E p_n &= V_n p_n + \epsilon_n^{-} f_n - p_{n-1} - p_{n+1} ,  \\
E f_n &= (E_{FB} + \epsilon_n^{+} ) f_n + \epsilon_n^{-} p_n,  
\end{eqnarray}
\label{eq:gen_fano_anderson} 
where $f_n$, $p_n$ are amplitudes of flat and dispersive band states at the $n$th unit cell of the lattice, disorder terms $V_n$ and $\epsilon_n^+$ account for intraband scattering within these bands, and $\epsilon^-$ describes the disorder scattering between the flat and dispersive bands. When the flat band is protected by a local symmetry, $\epsilon_n^+$ corresponds to the part of the disorder potential respecting this symmetry, and $\epsilon_n^{-}$ the antisymmetric part that breaks it (some explicit examples will be given below). Because of the quenched kinetic energy of the flat band, there is no direct hopping between the neighboring $f_n$. Therefore the flat band state amplitudes $f_n$ can be eliminated from Eqs.~(\ref{eq:gen_fano_anderson}), leaving
\begin{equation} 
E p_n = \left( V_n + \frac{(\epsilon_n^-)^2}{E - E_{FB} - \epsilon_n^+} \right) p_n - p_{n-1} - p_{n+1} . \label{eq:fano_eliminated}
\end{equation}
The second term in the brackets resembles a Fano resonance, depending on the detuning of the dispersive state from the energy of the CLS, $E-E_{FB}-\epsilon_n^+$. The effective onsite potential is resonantly enhanced, diverging when $E = E_{FB} + \epsilon_n^+$. Calculating the probability distribution function (PDF) of $z = 1/\epsilon_n^+$ yields~\cite{flach14}
\begin{equation*}
\mathcal{W}(z) = \frac{2}{z^2} \int \mathcal{P}(y) \mathcal{P}\left(\frac{2}{z} - y\right) dy,
\end{equation*}
where $\mathcal{P}(x)$ is the PDF of the random variable $\epsilon_n^+$. At $E = E_{FB}$, a bounded uniformly distributed disorder is transformed into one resembling a Cauchy distribution, with heavy tails $\mathcal{W}(z) \sim 1/z^2$ for large $z$. This type of distribution resembles the Lloyd model~\cite{lloyd1969} -- an Anderson-type model in which the onsite energies ${\tilde \epsilon}_n$ are drawn from a Cauchy distribution,
\begin{equation}
\mathcal{P}(x) = \frac{1}{\pi} \frac{W}{x^2 + W^2}. \label{eq:cauchy_distribution}
\end{equation}
The Lloyd model is exactly solvable. In particular, Thouless~\cite{thouless1972} and Ishii~\cite{ishii73} have derived its localization length $\xi$ in 1D, given as
\begin{equation}
4 \cosh ( \xi^{-1} ) = \sqrt{(2 + E)^2 + W^2} + \sqrt{(2 - E)^2 + W^2}.
\end{equation}
For energies within the band, $-2 < E < 2$, a Taylor expansion for small $\xi^{-1}$ and $W$ yields
\begin{equation}
\xi^{-1} = W / \sqrt{ 4 - E^2},
\end{equation}
Thus, $\xi \sim 1 / W^{\nu}$ displays a power law scaling with exponent $\nu = 1$, in contrast to the $\nu =2$ weak disorder scaling of the conventional 1D Anderson model~\cite{kappus81,derrida84}. This anomalous scaling occurs because the disorder distribution Eq.~\ref{eq:cauchy_distribution} has a divergent variance.

Similarly, at the dispersive band edge $E = \pm 2$, a weak disorder expansion yields
\begin{equation}
\xi^{-1} = \sqrt{2/W},
\end{equation}
with exponent $\nu = 1/2$, in contrast to the usual $\nu = 2/3$ when the disorder potential is bounded~\cite{derrida84,mielke85}. 

In other words, for small $W$ the Lloyd model displays much stronger localization than the regular Anderson model. This can be understood quite simply in terms of the divergent variance of the potential: given a long enough chain, one is certain to encounter a potential barrier large enough to classically trap the wave, \textit{i.e.} larger than the wave's kinetic energy. In contrast, the localization mechanism for small $W$ in the Anderson model is weak localization, an interference effect with no classical analogue, because the trapping potential is always smaller than the kinetic energy.

Based on this analogy, one can thus expect at $E = E_{FB}$ different power laws compared to the conventional Anderson model, originating from the Cauchy-tailed effective disorder potential. Heavy tails also persist for small detunings $|E - E_{FB} | = \delta$ from $E_{FB}$, as long as $\delta \lesssim W$ and $E - E_{FB} - \epsilon_n^+ = 0$ has a solution. For small shifts, one can thus expect a cross-over in scaling behaviour at $W \sim \delta$, \textit{e.g.} from $\nu = 1$ to $\nu = 2$.

When $E_{FB}$ is positioned in a gap, the dispersive states are evanescent even for vanishing disorder, and the distance from $E_{FB}$ to the dispersive band edge sets the
localization length similar to a defect state. Weak disorder merely determines the degree of hybridization between the evanescent waves and CLS. Hence we expect an exponent $\nu = 0$, with a nonzero finite $\xi ( W \rightarrow 0 )$ determined by the detuning of the flat band from the dispersive band edge.

Moving to higher dimensional lattices, $d > 1$, this Fano resonance picture will persist. It is, however, always possible for waves to travel around a resonant CLS state, so Anderson-localized dispersive waves cannot be trapped by Fano resonances and conventional scaling of $\xi$ is expected. On the other hand, the enhancement of intensity at resonance can still influence other measures of localization, such as the eigenmode participation number $P = 1/\sum_n |\Psi_n|^4$, which counts the number of strongly excited lattice sites.

All results from above have to be compared to the impact of the coupling of CLS amongst each other ($U>1$)  or to the potential coupling between
CLS and CLL. The network of weakly coupled and disordered CLS ($U>1$) will yield a localization length which is independent of the disorder strength
in the limit of weak disorder, since both the CLS-CLS coupling matrix elements and the CLS energy detuning are proportional to $W$. This effect will
therefore be of potential importance only for flat bands located in gaps, and the corresponding localization length has to be compared to the above discussed one which originates from the coupling between gapped CLS and dispersive states. As for the coupling between CLS and CLL, we have to refer to future work, as this aspect is currently not understood in detail.

\section{Examples}
\label{sec:examples}

We now apply the above formalism to some 1D examples generalizing the cross-stitch lattice studied in Ref.~\cite{flach14}. In particular, we consider the effect of onsite disorder on lattices with multiple flat bands (1D pyrochlore) and flat bands of class $U=2$ (stub and 1D Lieb)~\cite{flat_exciton_polariton,sawtooth}, and the effect of coupling disorder in the tunable diamond ladder~\cite{mukherjee2015}.

\subsection{1D Pyrochlore}

\begin{figure}

\includegraphics[width=\columnwidth]{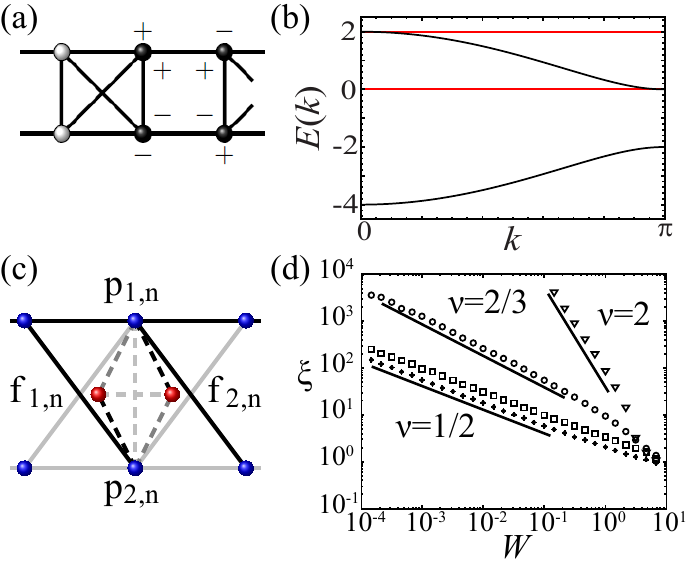}

\caption{(Color online) (a) The 1D pyrochlore lattice. Circles denote lattice sites, solid lines are hopping (off-diagonal) elements with strength 1. Filled circles show profile of a compact localized state, with alternating signs its amplitudes as indicated. (b) Irreducible band structure, for onsite energies $\epsilon = 0$. (c) Equivalent lattice detangled by Eq.~\ref{eq:1dpc_fano}. Solid lines indicate dispersive coupling terms (grey lines carry the opposite sign as compared to the black ones), while the dashed disorder-induced coupling terms are, from lightest to darkest, $\{\epsilon_n^{p2}, \epsilon_n^{f1}, \epsilon_n^{f1}\}$. (d) Scaling of Anderson localization length $\xi$ with disorder strength $W$, $\xi \sim 1/W^{\nu}$, for different energies $E$. Solid lines mark different power law exponents $\nu$. $E = -2$ (circles, $\nu = 2/3$, standard dispersive band edge), $E =0$ (squares, $\nu = 1/2$, flat band at dispersive band edge), $E=2$ (crosses, $\nu = 1/2$, flat band at dispersive band edge), $E=1$ (triangles, $\nu = 2$, standard dispersive band).}

\label{fig:1D-pyro}

\end{figure}

As an example of a $U=1$ flat band lattice, we consider the 1D pyrochlore lattice of Fig.~\ref{fig:1D-pyro}(a). Its amplitude equations read
\begin{eqnarray}
E \, a_n &= \epsilon_n^a a_n - b_n - b_{n-1} - d_n - c_{n-1}, \nonumber \\
E \, b_n &= \epsilon_n^b b_n - a_n - c_n - a_{n+1} - d_{n+1}, \nonumber \\
E \, c_n &= \epsilon_n^c c_n - b_n - d_n - a_{n+1} - d_{n+1}, \label{eq:1dpc} \\
E \, d_n &= \epsilon_n^d d_n - a_n - c_n - b_{n-1} - c_{n-1}. \nonumber
\end{eqnarray}
In the absence of the onsite potential $\epsilon$, there are two flat bands and two dispersive bands,
\begin{equation*}
E_{FB} = \left\lbrace 0,2 \right\rbrace, \quad E(k) = -1 \mp \sqrt{5 + 4 \cos k},
\end{equation*}
plotted in Fig.~\ref{fig:1D-pyro}(b). The first CLS ($E_{FB, 1}=0$) is
\begin{equation*}
a_n = c_n = \frac{1}{2}\delta_{n,m}, \quad b_n = d_n = -\frac{1}{2}\delta_{n,m} 
\end{equation*}
with a corresponding four-dimensional local basis vector of $\eclss{1} = (1,-1,1,-1)/2$. The second CLS ($E_{FB, 2}=2$) is
\begin{equation*}
a_n=b_n = \frac{1}{2}\delta_{n,m}, \quad c_n=d_n = -\frac{1}{2}\delta_{n,m}
\end{equation*}
with a local basis vector of $\eclss{2}= (1, 1, -1,-1)/2$. These two local basis vectors define two decoupled flat band states $f_{\{1,2\},n}$ for each unit cell,
\begin{eqnarray*}
f_{\{1,2\},n} &= \frac{1}{2}(a_n \mp b_n \pm c_n-d_n), \\ 
\epsilon^{f_{\{1,2\}}}_{n} &= \frac{1}{4}(\epsilon^a_n \mp \epsilon^b_n \pm \epsilon^c_n-\epsilon^d_n) 
\end{eqnarray*}
while the remaining two-dimensional orthogonal subspace spanned by $\ep$ encodes the two remaining dispersive degrees of freedom $p_{\{1,2\},n}$, corresponding to the two dispersive spectral branches of Fig.~\ref{fig:1D-pyro}(b),
\begin{eqnarray*}
p_{\{1,2\},n} &= \frac{1}{2}(a_n \pm b_n \pm c_n+d_n), \\ 
\epsilon^{p_{\{1,2\}}}_{n} &= \frac{1}{4}(\epsilon^a_n \pm \epsilon^b_n \pm \epsilon^c_n+\epsilon^d_n).
\end{eqnarray*}
These transformations can be encoded in the local rotation $\hat{D}$,
\begin{equation}
\hat{D} = \frac{1}{2} \left( \begin{array}{cccc} 1 & 1 & 1 & 1 \\ 1 & -1 & -1 & 1 \\ 1 & -1 & 1 & - 1 \\ 1 & 1 & -1 & -1 \end{array} \right),
\end{equation}
with $(p_{1,n},p_{2,n},f_{1,n},f_{2,n} ) = \hat{D} (a_n,b_n,c_n,d_n)$, and the diagonal disorder terms transforming to $\hat{D} \mathrm{diag}(\epsilon_n^j) \hat{D}^{-1}$ (here $\mathrm{diag}(\epsilon_n^j)$ denotes a diagonal matrix with elements $\epsilon_n^j$ on the diagonal). The transformed disorder thus has both diagonal and off-diagonal components. Rewriting Eq.~\ref{eq:1dpc} in terms of the detangled variables, we obtain an equivalent set of amplitude equations,
\begin{eqnarray}
E f_{1,n} &= \left( \epsilon^{p_1}_n + 2\right) f_{1,n} + \epsilon^{p_2}_n f_{2,n} + \epsilon^{f_1}_n p_{1,n} + \epsilon^{f_2}_n p_{2,n}, \nonumber \\
E f_{2,n} &= \epsilon^{p_2}_n f_{1,n} + \epsilon^{p_1}_n f_{2,n} + \epsilon^{f_2}_n p_{1,n} + \epsilon^{f_1}_n p_{2,n}, \label{eq:1dpc_fano} \\
E p_{1,n} &= \epsilon^{f_1}_n f_{1,n} + \epsilon^{f_2}_n f_{2,n} + \left(\epsilon^{p_1}_n - 2\right) p_{1,n} + \epsilon^{p_2}_n p_{2,n} - \nonumber \\
& p_{1,n-1}  - p_{1,n+1} + p_{2,n-1}  - p_{2,n+1}, \nonumber \\
E p_{2,n} &= \epsilon^{f_2}_n f_{1,n} + \epsilon^{f_1}_n f_{2,n} + \epsilon^{p_2}_n p_{1,n} + \epsilon^{p_1}_n p_{2,n} + \nonumber \\
& p_{2,n-1}  + p_{2,n+1}  - p_{1,n-1}  + p_{1,n+1}, \nonumber
\end{eqnarray}
corresponding to the ``detangled'' partially diagonalized lattice in Fig.~\ref{fig:1D-pyro}(c). Note in particular the structure of the effective coupling terms: symmetric parts yielding effective onsite potentials, and antisymmetric parts locally coupling the flat band states to the dispersive degrees of freedom, exactly as advertised in the previous section.

Hence if the potential satisfies the local symmetry 
$\epsilon^{f1}_n = \epsilon^{f2}_n=0$ then the two CLS $f_{\{1,2\},n}$ decouple from the dispersive degrees of freedom $p_{\{1,2\},n}$. 
If that symmetry is additionally supported on all unit cells, then all CLS decouple with the individual energies
\begin{equation*}
 E=1+\epsilon^{p1}_n \pm \sqrt{1+\left(\epsilon^{p2}_n\right)^2}.
\end{equation*}
Note that if $\epsilon^{p2}_n \neq 0$, the two CLS hybridize, but retain their compactness otherwise.  

Because the coupling to the CLS is purely \emph{local}, one can use the first two lines of Eq.~\ref{eq:1dpc_fano} to eliminate the $f_{\{1,2\},n}$ variables for \emph{arbitrary} $\epsilon_n$, leaving equations for the dispersive variables $p_{\{1,2\},n}$ subject to a resonantly-enhanced effective disorder potential, in the same manner as the previous section. We hence expect quantities such as the localization length to be modified by an effective heavy-tailed disorder following a Cauchy distribution. 

We verify this by numerically calculating the localization length $\xi( W )$ using the transfer number method outlined in the Appendix, assuming uncorrelated and uniformly distributed disorder $\epsilon \in [-W/2,W/2]$, where $W$ is the disorder strength. We consider the scaling of $\xi$ with $W$ at different energies in Fig.~\ref{fig:1D-pyro}(d). Away from the flat bands, we observe the usual $\nu = 2$ and $\nu = 2/3$ within and at band edges respectively, while at each of the flat band energies an anomalous $\nu = 1/2$ is observed. Thus, modelling the flat band as a set of decoupled CLS remains a good approximation in lattices with multiple flat bands, even when disorder introduces (weak) hybridization between their CLS.

\subsection{1D Stub}

\begin{figure}

\includegraphics[width=\columnwidth]{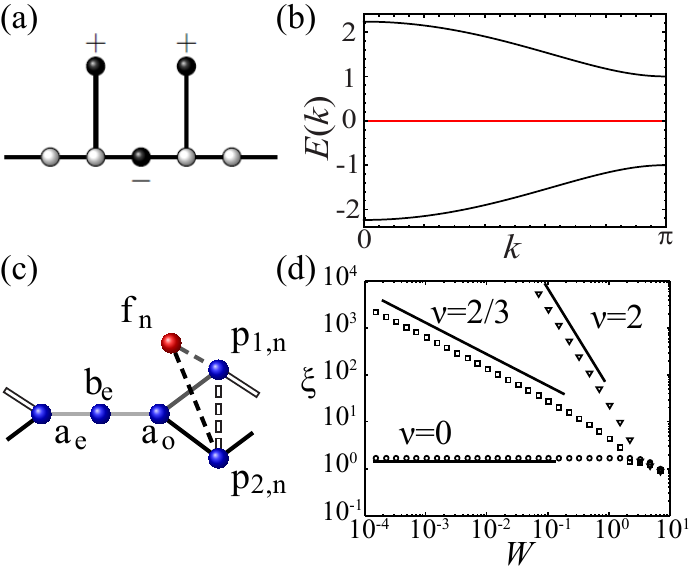}

\caption{(Color online) (a) The stub lattice. Circles denote lattice sites, solid lines are hopping (off-diagonal) elements with strength 1. Filled circles show profile of a compact localized state, with alternating signs its amplitudes as indicated. (b) Irreducible band structure, for onsite energies $\epsilon = 0$. (c) Equivalent lattice detangled by Eq.~\ref{eq:stub_fano}. Dispersive couplings are $\pm\sqrt{1/2}, -1, -\sqrt{3/2}$ (from light to dark). The disorder-induced couplings are respectively $\epsilon_n/\sqrt{2}, \epsilon_n, \gamma_n$ (from light to dark). To conserve space, the notations $a_e = a_{2n-2}, b_e = b_{2n-2}, a_o = a_{2n-1}$ have been used. (d) Scaling of Anderson localization length $\xi$ with disorder strength $W$, $\xi \sim 1/W^{\nu}$, for different energies $E$. Solid lines mark different power law exponents $\nu$. $E = 0$ (circles, $\nu = 0$, gapped flat band), $E =1$ (squares, $\nu = 2/3$, standard dispersive band edge), $\sqrt{2}$ (triangles, $\nu = 2$, standard dispersive band).}

\label{fig:1D-stub}

\end{figure}

For our second example we consider the 1D stub lattice shown in Fig.~\ref{fig:1D-stub}(a), which is of class $U=2$. Its eigenmode equations are
\begin{eqnarray}
E a_n &= \epsilon_n^a a_n - b_{n-1} -b_n - c_n, \nonumber \\
E b_n &= \epsilon_n^b b_n - a_{n} - a_{n+1}, \label{eq:stub} \\
E c_n &= \epsilon_n^c c_n - a_n \nonumber ,
\end{eqnarray}
with spectrum (for $\epsilon=0$)
\begin{equation*}
E_{FB} = 0, \quad E(k) = \pm \sqrt{3 + 2 \cos k},
\end{equation*}
plotted in Fig.~\ref{fig:1D-stub}(b). A CLS occupies a minumum of two unit cells,
\begin{equation*} 
a_n = 0, \; b_n = -\frac{1}{\sqrt{3}}\delta_{n,m}, \; c_n = \frac{1}{\sqrt{3}} ( \delta_{n,m} + \delta_{n,m+1} ).
\end{equation*}
To detangle the lattice, we introduce a supercell formed by two unit cells, and perform rotations in the 3-dimensional subspace occupied by the CLS (leaving the $a_n$ sublattice invariant). This detangling will only account for half of the flat band states, and we wish to determine whether this partial detangling is sufficient to understand the properties of the disordered system. The transformed amplitudes
\begin{eqnarray}
f_n &= \frac{1}{\sqrt{3}} (c_{2n-1} - b_{2n-1} + c_{2n}), \\
p_{1,n} &= \frac{1}{\sqrt{2}} ( c_{2n-1} - c_{2n} ), \\
p_{2,n} &= \frac{1}{\sqrt{6}} ( c_{2n-1} + 2 b_{2n-1} + c_{2n} ),
\end{eqnarray}
correspond to the local rotation $(p_{1,n},p_{2,n},f_n) = \hat{D}(c_{2n-1},b_{2n-1},c_{2n})$ with
\begin{equation}
\hat{D} = \left( \begin{array}{ccc} \frac{1}{\sqrt{2}} & 0 & -\frac{1}{\sqrt{2}} \\ \frac{1}{\sqrt{6}} & \frac{2}{\sqrt{6}} & \frac{1}{\sqrt{6}} \\ \frac{1}{\sqrt{3}} & -\frac{1}{\sqrt{3}} & \frac{1}{\sqrt{3}} \end{array} \right).
\end{equation}
Similarly transforming the potentials,
\begin{eqnarray*}
\epsilon_n^f &= \frac{1}{3}(\epsilon_{2n-1}^c + \epsilon_{2n-1}^b + \epsilon_{2n}^c ), \\
\epsilon_n &= \frac{1}{\sqrt{6}} ( \epsilon_{2n-1}^c - \epsilon_{2n}^c ), \\
\gamma_n &= \frac{1}{3\sqrt{2}} ( \epsilon_{2n-1}^c - 2 \epsilon_{2n-1}^b + \epsilon_{2n}^c ),
\end{eqnarray*}
and substituting into Eq.~\ref{eq:stub}, we obtain the equivalent detangled lattice
\begin{eqnarray}
E f_n &=& \epsilon_n^f f_n + \epsilon_n \, p_{1,n} + \gamma_n \, p_{2,n}, \nonumber \\
E p_{1,n} &=&  \epsilon_n f_n + \left(\epsilon_n^f +\frac{\gamma_n}{\sqrt{2}}\right) p_{1,n} + \frac{\epsilon_n}{\sqrt{2}}\, p_{2,n} \nonumber \\
&& + \frac{1}{\sqrt{2}} \left(a_{2n} - a_{2n-1}\right), \nonumber \\
E p_{2,n} &=& \gamma_n f_n + \frac{\epsilon_n}{\sqrt{2}}\, p_{1,n} + \left(\epsilon_n^f -\frac{\gamma_n}{\sqrt{2}}\right) p_{2,n} \nonumber \\
&& - \sqrt{\frac{3}{2}} \left(a_{2n-1} + a_{2n} \right), \label{eq:stub_fano} \\
E a_{2n-1} &=& \epsilon^a_{2n-1} a_{2n-1} - b_{2n-2} -\frac{1}{\sqrt{2}} p_{1,n} - \sqrt{\frac{3}{2}} p_{2,n}, \nonumber \\
E a_{2n} &=& \epsilon^a_{2n} a_{2n} - b_{2n} +\frac{1}{\sqrt{2}} p_{1,n} - \sqrt{\frac{3}{2}} p_{2,n}, \nonumber \\
E b_{2n} &=& \epsilon^b_{2n} b_{2n} - a_{2n} - a_{2n+1}, \nonumber
\end{eqnarray}
illustrated in Fig.~\ref{fig:1D-stub}(c). The $p_{1,j},p_{2,j}$ sites form two separate paths between the $a_{2j-1}$ and $a_{2j}$ sites; destructive interference between these paths can suppress wave transport, corresponding to the ``missing'' flat band states not included in our detangling. In fact, these missing states form a flat band of class $U=2$ in the detangled lattice; thus by repeatedly applying the detangling partial diagonalization one can isolate a larger fraction of the FBS to identify symmetries in the disorder potential that leave them decoupled. For example, the detangled flat band states $f_j$ are completely decoupled from the chain in the absence of disorder, and they remain decoupled if the disorder has the local symmetry $\epsilon_n = \gamma_n = 0$.

Turning to the scaling of the localization length in Fig.~\ref{fig:1D-stub}(d), we observe the expected exponents $\nu = 2/3$ and $2$ for the dispersive bands. Since the flat band is isolated from the dispersive bands by a gap, the localization length at $E_{FB}$ remains finite and independent of $W$ as $W \rightarrow 0$, i.e. $\nu = 0$; the quenched kinetic energy of the flat band means there is no energy scale to compete with $W$ in the weak disorder limit. Hence $\xi \approx 1.692$ remains finite as $W \rightarrow 0$. Interestingly the estimation of $\xi$ in Sec.~\ref{sec:disorder} based on solving $E_{FB} = E(k)$ for $k$ to obtain the evanescent decay rate of the dispersive states at $E_{FB}$ predicts stronger localization, $\xi = 1.04$. This implies that the disorder-induced hybridization between the CLS in $U>1$ lattices is increasing the localization length compared to the $U=1$ case. 

For small energy shifts $\delta E$ from $E_{FB}$, a crossover from disorder-induced localization to band gap-induced localization will occur when $W \lesssim \delta E$, with $\xi$ remaining finite. This crossover is responsible for the nontrivial behaviour of $\xi(E,W)$ recently reported in Ref.~\cite{ge_arxiv}.

\subsection{1D Lieb}

\begin{figure}

\includegraphics[width=\columnwidth]{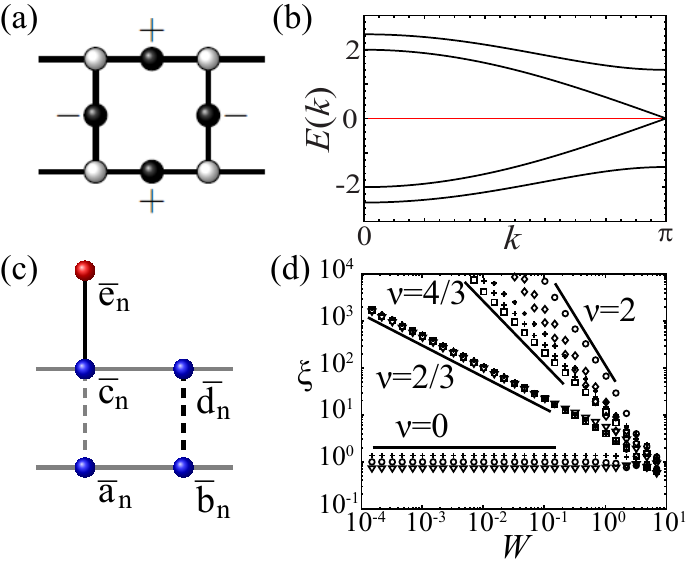}

\caption{(Color online) (a) The 1D Lieb lattice. Circles denote lattice sites, solid lines are hopping (off-diagonal) elements with strength 1. Filled circles show profile of a compact localized state, with alternating signs its amplitudes as indicated. (b) Irreducible band structure, for onsite energies $\epsilon = 0$. (c) Equivalent lattice detangled by Eq.~\ref{eq:lieb_fano}.  The fixed couplings are $-1,-\sqrt{2}$ (light/dark). The disorder-induced couplings are $\epsilon_n^{1-}$ (light) and $\epsilon_n^{2-}$ (dark) (d) Scaling of Anderson localization length $\xi$ with disorder strength $W$, $\xi \sim 1/W^{\nu}$, for different energies $E$. Solid lines mark different power law exponents $\nu$. $E = 0$ (crosses, $\nu = 0,4/3$), $E =1$ (circles, $\nu = 0,2$), $\sqrt{2}$ (points, $\nu = 2/3,4/3$), $E =1.8$ (diamonds, $\nu = 2,2$), $E =2$ (squares, $\nu = 2/3,4/3$), $E =\sqrt{6}$ (triangles, $\nu = 0,2/3$).
}

\label{fig:1D_lieb}

\end{figure}

As a second example of class $U=2$ we consider the 1D Lieb lattice shown in Fig.~\ref{fig:1D_lieb}(a). Its eigenmode equations
\begin{eqnarray}
E a_n &= \epsilon_{n}^a a_n - d_{n-1} - d_n - e_n, \nonumber \\
E b_n &= \epsilon_{n}^b b_n - c_{n-1} - c_n - e_n, \nonumber \\
E c_n &= \epsilon_{n}^c c_n - b_{n} - b_{n+1}, \label{eq:1D_lieb} \\
E d_n &= \epsilon_{n}^d d_n - a_n - a_{n+1}, \nonumber \\
E e_n &= \epsilon_{n}^e e_n - a_n - b_n, \nonumber
\end{eqnarray}
yield (for $\epsilon=0$) the spectrum 
\begin{equation*}
E_{FB} = 0, \; E(k) = \left\lbrace \pm \sqrt{2 + 2 \cos k}, \pm \sqrt{4 + 2 \cos k} \right\rbrace,
\end{equation*}
plotted in Fig.~\ref{fig:1D_lieb}(b). In contrast to the stub lattice, the Lieb lattice has dispersive bands in resonance with its flat band, so we anticipate nontrivial scaling of its localization length. The flat band's CLS is
\begin{equation*}
a_n = b_n = 0, \; c_n = d_n = \frac{1}{2} \delta_{n,m}, \; e_n = -\frac{1}{2} (\delta_{n,m} + \delta_{n,m+1} ). 
\end{equation*}
Applying the transformation
\begin{equation*}
\left\lbrace \bar{a}_n, \bar{c}_n \right\rbrace = \frac{1}{\sqrt{2}} \left( a_n \mp b_n \right), \quad
\left\lbrace \bar{b}_n, \bar{d}_n \right \rbrace = \frac{1}{\sqrt{2}} \left( d_n \mp c_n \right) 
\end{equation*}
corresponding to the rotation
\begin{equation*}
\hat{D} = \frac{1}{\sqrt{2}} \left( \begin{array}{ccccc} 1 & -1 & 0 & 0 & 0 \\ 0 & 0 & -1 & 1 & 0 \\ 1 & 1 & 0 & 0 & 0 \\ 0 & 0 & 1 & 1 & 0 \\ 0 & 0 & 0 & 0 & \sqrt{2} \end{array} \right), 
\end{equation*}
and similarly transforming the disorder potential by $\hat{D} \mathrm{diag}( \epsilon_{n}^j ) \hat{D}^{-1}$,
\begin{equation*}
\epsilon_n^{1\pm} = \frac{1}{2} \left( \epsilon_n^a \pm \epsilon_n^b \right), \quad
\epsilon_n^{2\pm} = \frac{1}{2} \left( \pm \epsilon_n^c + \epsilon_n^d \right),
\end{equation*}
we obtain the partially-detangled lattice
\begin{eqnarray}
E \bar{a}_n &= \epsilon_n^{1+} \bar{a}_n - \bar{b}_{n-1} - \bar{b}_n + \epsilon_n^{1-} \bar{c}_n, \nonumber \\
E \bar{b}_n &= \epsilon_n^{2+} \bar{b}_n - \bar{a}_n - \bar{a}_{n+1} + \epsilon_n^{2-} \bar{d}_n, \nonumber \\
E \bar{c}_n &= \epsilon_n^{1+} \bar{c}_n + \epsilon_n^{1-} \bar{a}_n - \bar{d}_{n-1} - \bar{d}_n -  e_n \sqrt{2},  \label{eq:lieb_fano} \\
E \bar{d}_n &= \epsilon_n^{2+} \bar{d}_n + \epsilon_n^{2-} \bar{b}_n - \bar{c}_n - \bar{c}_{n+1}, \nonumber  \\
E e_n &= \epsilon_n^e e_n - \bar{c}_n \sqrt{2}, \nonumber 
\end{eqnarray}
shown in Fig.~\ref{fig:1D_lieb}(c). It consists of an ordinary 1D chain with ($\bar{a}_n,\bar{b}_n$) sites, corresponding to the middle two dispersive bands, coupled via disorder to a 1D stub lattice. One could further detangle this stub lattice is a manner similar to the calculations above, but this partial detangling is already sufficient to understand the effect of disorder.

In contrast to the previous examples, in the Lieb lattice there are two possible transmission channels for the dispersive waves, corresponding to either $(\bar{a}_n,\bar{b}_n)$ or $(\bar{c}_n,\bar{d}_n)$ chains, leading to two distinct localization lengths which can be computed using the transfer matrix method outlined in the Appendix. The observed power law scalings in Fig.~\ref{fig:1D_lieb}(d) show some interesting features caused by either the resonant interaction between CLS and a dispersive channel, or between the two dispersive channels.

At $E_{FB}$ there are two exponents: $\nu = 0$, and $\nu \approx 4/3$. The $\nu = 0$ exponent corresponds to transmission along the ``stub'' part of the detangled lattice; since its flat band is separated by a gap, there is no scaling of its localization length, as expected. On the other hand, modes of the $(\bar{a}_n,\bar{b}_n)$ chain are in resonance with the stub's flat band states, leading to a Fano resonance-induced modification to its scaling. Since the CLS do not excite the $\bar{c}_n$ sublattice, disorder can only directly couple the flat band states to the $\bar{b}_n$ sublattice, \textit{i.e.} there is only a Fano resonance at every \emph{second} lattice site. This resembles the geometry of the diamond ladder studied extensively in Ref.~\cite{fb0}, and remarkably we obtain the same scaling exponent $\nu \approx 4/3$ even though the physical lattice is completely different.

A second pair of anomalies occur at the energies $E = \sqrt{2},2$, which both have exponents $\nu = 2/3,4/3$. At these energies one of the dispersive channels has a band edge, corresponding to exponent $\nu = 2/3$. Interestingly, the vanishing group velocity and stronger localization associated with these band edge states presumably leads to the resonantly-enhanced $\nu = 4/3$ exponent for the mid-band states in the other dispersive channel (a full analysis of the resonant interaction between multiple dispersive channels is however beyond the scope of this work). All other energies display conventional exponents.

\subsection{1D Tunable Diamond: Off-diagonal disorder}

\begin{figure}
\centering
\includegraphics[width=\columnwidth]{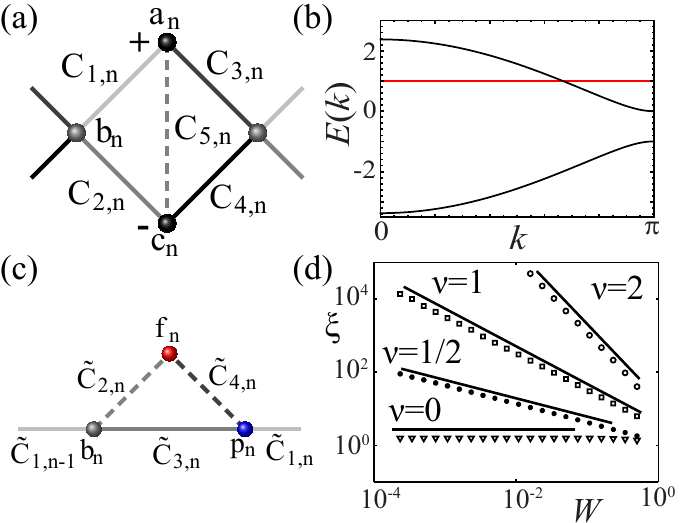}
\caption{(Color online) (a) The 1D diamond ladder. Circles denote lattice sites, solid lines are hopping (off-diagonal) elements with random strength. Filled circles show profile of a compact localized state, with alternating signs its amplitudes as indicated. (b) Irreducible band structure, for $t=1$ and onsite energies $\epsilon = 0$.
(c) Equivalent lattice detangled by Eq.~\ref{eq:diamond_offd_fano}. (d) Scaling of Anderson localization length $\xi$ with disorder strength $W$, $\xi \sim 1/W^{\nu}$, for different energies $E$. Solid lines mark different power law exponents $\nu$. $E=1.5, t=1$ (open circles, $\nu = 2$, standard dispersive band), $E=t=1$ (squares, $\nu = 1$, flat band inside dispersive band), $E=t=2$ (filled circles, $\nu = 1/2$, flat band at dispersive band edge), $E = t = 2.1$ (triangles, $\nu = 0$, gapped flat band).}
\label{fig:diamond_offd}
\end{figure}

As a final application of the partial diagonalization ``detangling'' procedure in 1D, we consider off-diagonal disorder~\cite{femto_offdiagonal}. In bipartite lattices, off-diagonal disorder has a significantly different effect compared to diagonal disorder: modes at the band centre $E = 0$ become delocalized \cite{offdiagonal_theory,offdiag2,offdiag3}.
However in our case, the detangling change of basis generically mixes disorder terms into a combination of effective onsite and hopping disorder. Thus, pure hopping disorder and pure onsite disorder have qualitatively similar effects, so there are no delocalized states at $E_{FB}$. As an explicit example, consider the $U=1$ diamond ladder with hopping disorder shown in Fig.~\ref{fig:diamond_offd}(a). Its eigenmode equations read
\begin{eqnarray}
E a_n &=& -C_{1,n} b_n - C_{3,n} b_{n+1} - C_{5,n} c_n, \nonumber \\
E b_n &=& -C_{3,n-1} a_{n-1} - C_{1,n} a_n - C_{4,n-1} c_{n-1} - C_{2,n} c_n, \nonumber \\
E c_n &=& -C_{2,n} b_n - C_{4,n} b_{n+1} - C_{5,n} a_n, \label{eq:diamond_offd}
\end{eqnarray}
where the hopping is taken from uniformly distributed random variables $C_{1-4,n} \in 1 + [- W/2, W/2]$, and $C_{5,n} \in t + [-W/2, W/2]$. To ensure positive nonzero couplings -- generally the case for evanescently coupled waveguide arrays -- we enforce the restriction $W < \lfloor t, 1 \rfloor$. In the ordered case $W=0$ the cross coupling term $C_{5,n}$ shifts the energy of the flat band to $E_{FB} = t$, opening a gap between the two dispersive bands, see eg. Fig.~\ref{fig:diamond_offd}(b) showing the spectrum for $t=1$. When $t>2$, $E_{FB}$ becomes separated from dispersive bands by a gap.

We reduce the eigenvalue problem to its detangled form via the change of basis
\begin{equation*}
f_n = \frac{C_{4,n} a_n - C_{3,n} c_n }{ \sqrt{C_{3,n}^2 + C_{4,n}^2 }}, \quad p_n = \frac{C_{3,n} a_n + C_{4,n} c_n}{\sqrt{C_{3,n}^2 + C_{4,n}^2}},
\end{equation*}
corresponding to $(p_n,f_n) = \hat{D} (a_n,c_n)$ with
\begin{equation}
\hat{D} = \frac{1}{\sqrt{C_{3,n}^2 + C_{4,n}^2}} \left( \begin{array}{cc} C_{4,n} & -C_{3,n} \\ C_{3,n} & C_{4,n} \end{array} \right).
\end{equation}
Notably, here the change of basis itself depends on the disorder. The eigenmode equations become
\begin{eqnarray}
(E - C_{5,n} ) f_n &= -\tilde{C}_{2,n} b_n - \tilde{C}_{4,n} p_n, \nonumber \\
(E + C_{5,n} ) p_n &= -\tilde{C}_{3,n} b_n - \tilde{C}_{1,n} b_{n+1} - \tilde{C}_{4,n} f_n, \label{eq:diamond_offd_fano} \\
E b_n &= -\tilde{C}_{2,n} f_n - \tilde{C}_{3,n} p_n - \tilde{C}_{1,n-1} p_{n-1}, \nonumber
\end{eqnarray}
where we have introduced the effective couplings 
\begin{eqnarray}
\tilde{C}_{1,n} &= \sqrt{ C_{3,n}^2 + C_{4,n}^2 },\nonumber \\
\tilde{C}_{2,n} &= (C_{1,n} C_{4,n} - C_{2,n} C_{3,n} )/\tilde{C}_{1,n},\nonumber \\
\tilde{C}_{3,n} &= (C_{1,n} C_{3,n} + C_{2,n} C_{4,n} )/\tilde{C}_{1,n}, \\
\tilde{C}_{4,n} &= C_{5,n} (C_{4,n}^2 - C_{3,n}^2) / (2 C_{3,n} C_{4,n} ). \nonumber
\end{eqnarray}
This equivalent lattice is illustrated in Fig.\ref{fig:diamond_offd}(c). The off-diagonal disorder is transformed to a mixture of effective on-site and hopping disorder. The flat band states have energy $C_{5,n}$ and are coupled to \emph{both} sublattices of the dispersive chain. In the limit of vanishing disorder, $\tilde{C}_{2,n}, \tilde{C}_{4,n} \rightarrow 0$. Then $\tilde{C}_{1,n}, \tilde{C}_{3,n} \rightarrow \sqrt{2}$ and coupling to the flat band states vanishes. Such decoupling also occurs if the disorder is correlated such that $C_{1,n} = C_{2,n}$ and $C_{3,n} = C_{4,n}$.

Eliminating the flat band states $f_n$, we obtain
\begin{eqnarray}
E p_n &= \epsilon_{1,n} p_n - \tilde{C}_{0,n} b_n - \tilde{C}_{1,n} b_{n+1}, \nonumber  \\
E b_n &= \epsilon_{2,n} b_n - \tilde{C}_{0,n} p_n - \tilde{C}_{1,n-1} p_{n-1}, \nonumber 
\end{eqnarray}
where
\begin{eqnarray*}
\epsilon_{1,n} \equiv &  \frac{\tilde{C}_{4,n}^2}{E + C_{5,n}} + C_{5,n}, \quad \epsilon_{2,n} \equiv \frac{\tilde{C}_{2,n}^2}{E + C_{5,n}}, \\
& \tilde{C}_{0,n}  \equiv  \tilde{C}_{3,n} + \frac{\tilde{C}_{2,n} \tilde{C}_{4,n} }{E + C_{5,n}},
\end{eqnarray*}
which is similar to the onsite disorder case, as expected. In particular, for weak disorder we have $\tilde{C}_{2,n},\tilde{C}_{4,n} \sim W$ and the Fano resonance leads to a divergence of the effective onsite potentials $\varepsilon_{j,n}$ when $E = C_{5,n}$. In the disordered system this can occur within the resonance width $|E - E_{FB} | < W/2$, leading again to an effective disorder potential with Cauchy tails and a much smaller localization length. Outside this window, the effective onsite disorder is bounded with (nonzero) strength proportional to $W$.

As above, we test these predictions by numerically calculating $\xi(E,W)$ using the transfer number method. We find indeed that $\xi (E)$ shows no divergences, and calculate the weak disorder power law scalings at $E_{FB}$ for different flat band positions (choices of $t$) in Fig.~\ref{fig:diamond_offd}. In all cases, the predicted anomalous scaling laws at $E_{FB}$ are observed ($\nu = 1,1/2$ or 0), in contrast to the $\nu = 2$ off-resonance with the flat band. Thus, as anticipated, hopping disorder has an effect qualitatively similar to onsite disorder.

\section{2D Lattices}
\label{2d}

\begin{figure}
\centering
\includegraphics[width=\columnwidth]{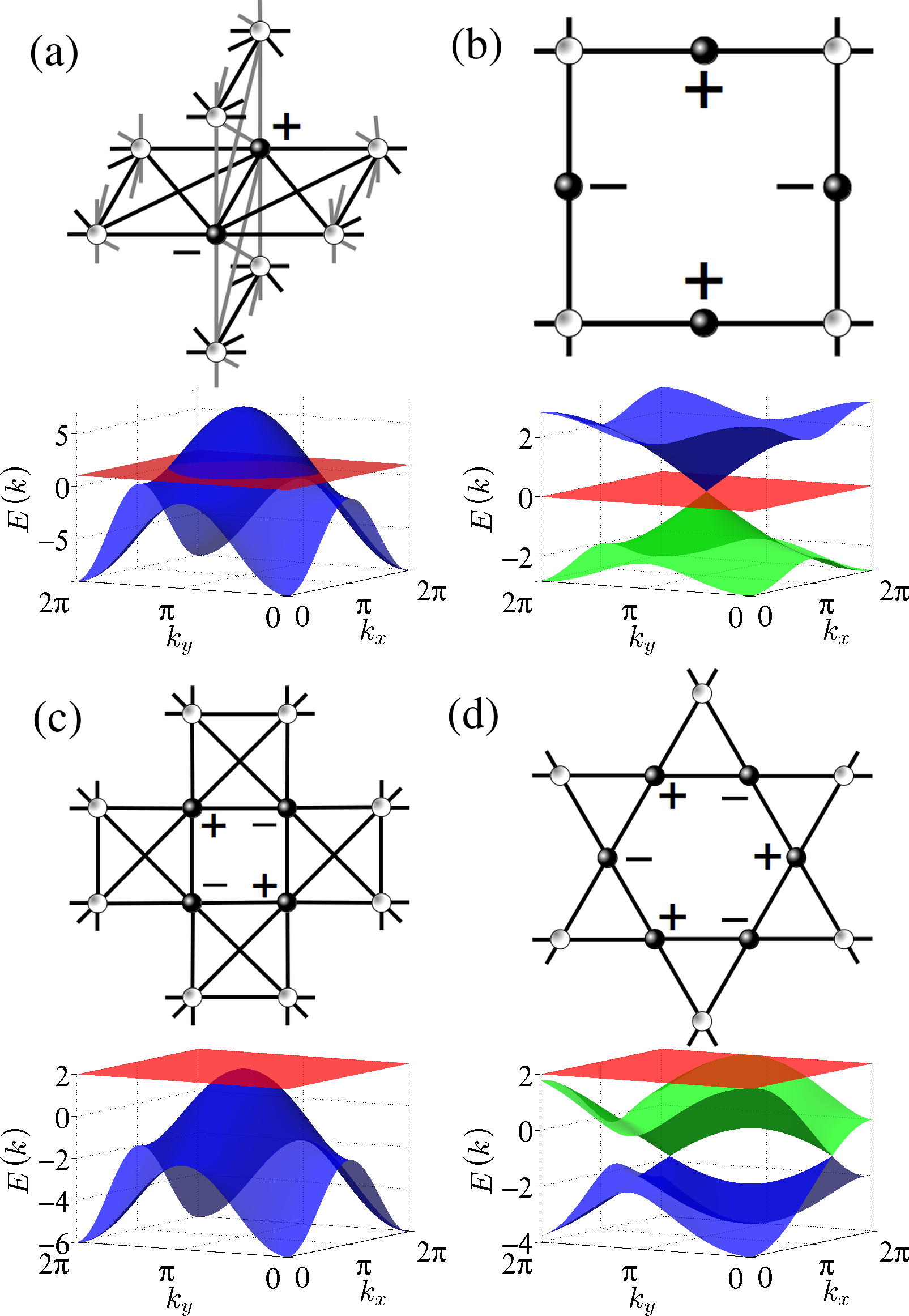}
\caption{(Color online) Two-dimensional FB lattices. Circles denote lattice sites, black/grey lines are hopping (off-diagonal) elements of $t=1$. Filled circles again show CLS location. Irreducible band structures for onsite energies $\epsilon=0$ are shown below each lattice, with flat bands corresponding to red horizontal planes. The models and CLS classes are: (a) 2-D cross-stitch, $U=1$ (black/grey lines denote horizontal/vertical bonding); (b) Lieb, $U=3$; (c) pyrochlore, $U=3$; and (d) Kagome, $U=4$.   }
\label{fig2}
\end{figure}

We now examine how the detangling partial diagonalization generalizes to 2D lattices. Fig.~\ref{fig2} shows examples of 2D flat band lattices along with their CLS and irreducible band structures. For $U=1$ lattices such as the cross-stitch, the position of the flat band with respect to dispersive bands is tunable, whereas in the $U>1$ examples (2D Lieb, 2D pyrochlore, and Kagome) the flat band is forced to touch the edge of a dispersive band~\cite{compact_lieb_1,compact_lieb_2,induced_kagome,induced_lieb}. These ``protected'' crossings are a consequence of the real space topology of the CLS, which results in an overcomplete basis of flat band states; there is no way to introduce a band gap without destroying the CLS and unflattening the band~\cite{bergman2008}.

As noted above, in dimensions $d > 1$ it is always possible for a dispersive wave to travel around a strongly scattering localized flat band state, which spoils the trapping mechanism responsible for the anomalous scaling of the localization length observed in the 1D examples. We thus expect on one hand conventional scaling of the localization length $\xi$  in these 2D examples, \textit{i.e.} marginal localization with an essential singularity as $W \rightarrow 0$~\cite{kramer93}.

On the other hand, resonances still result in very strongly excited sites, which influences other measures of eigenmode localization such as the participation number $P = 1/\sum_n |\Psi_n|^4$. $P$ counts the number of sites strongly excited in an eigenmode $\Psi_n$, and in conventional lattices it is proportional to the localization volume $(\xi)^d$. In a weakly disordered flat band, $P$ should instead count the number of resonant flat band states, $\sim W (\xi)^d$, significantly smaller than the localization volume in the weak disorder limit $W \ll 1$~\cite{flach14}. 

\begin{figure}

\includegraphics[width=\columnwidth]{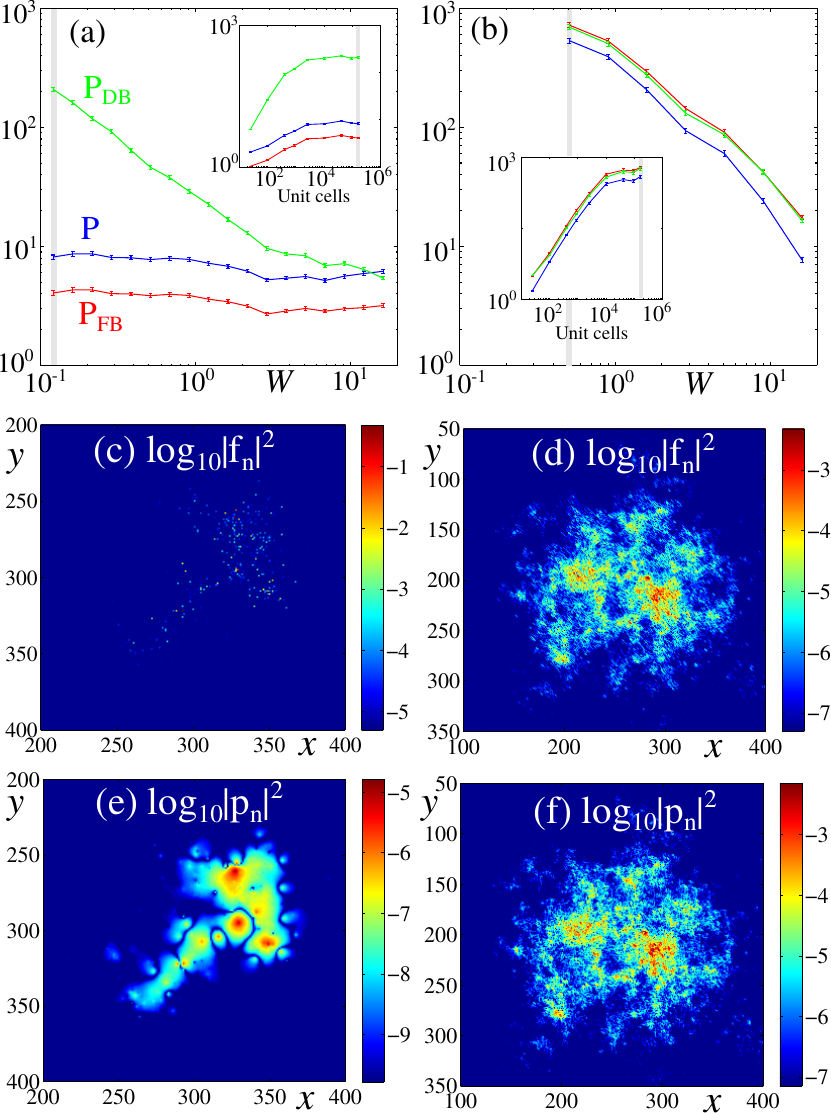}

\caption{(Color online) Flat band Anderson localization in the 2D cross-stitch (a,c,e) and pyrochlore (b,d,f) lattices. (a,b) Scaling of mean eigenmode participation number $P$ with disorder strength $W$ at $E_{FB}$, along with its flat band $P_{\mathrm{FB}}$ and dispersive components $P_{\mathrm{DB}}$. The lattice size is $N_x\times N_y = 400\times400$ unit cells (periodic boundary conditions) and error bars indicate statistical error. Insets show finite size scaling of the results at $W_{\mathrm{min}}$ (gray vertical bar), illustrating convergence to the large $N$ limit. (c--f) Representative intensity profiles of flat (c,d) and dispersive (e,f) band components of a weak disorder eigenmode at $W_{\mathrm{min}}$.}

\label{fig:2D_scaling}

\end{figure}

To test this prediction, we carry out exact diagonlization of disordered $t=4$ cross-stitch and pyrochlore lattices. In both cases the flat band touches the dispersive band edge. For eigenmodes near the flat band energy $E_{FB}$, we compute three measures: $P$, $P_{\mathrm{FB}}$, and $P_{\mathrm{DB}}$. $P_{\mathrm{FB}}$ is the participation number of the normalized flat band component of the eigenmode, estimating the number of resonant CLS. Similarly the participation number of the dispersive band component $P_{\mathrm{DB}}$ estimates the localization volume, $P_{\mathrm{DB}} \sim (\xi)^d$. Fig.~\ref{fig:2D_scaling} shows the scaling of these three measures with disorder strength $W$. 

For the cross-stitch case we indeed see that $P$ counts the number of resonances $\sim P_{FB}$ rather than the localization volume $\sim P_{\mathrm{DB}}$, which grows much more rapidly as $W \rightarrow 0$. The slower growth of $P_{\mathrm{FB}}$ is consistent with the number of resonances growing as $P_{\mathrm{FB}} \propto W (\xi)^d$. On the other hand, in the 2D pyrochlore we see a distinctly different behaviour, with $P \sim P_{\mathrm{FB}} \sim P_{\mathrm{DB}}$ all growing at similar rates as $W \rightarrow 0$. Thus, the predicted scaling of the number of resonances in an eigenmode breaks down at a protected crossing.

To further demonstrate this strong qualitative difference, we consider the profiles of the flat band and dispersive band components of a representative eigenmode at $E_{FB}$ in Fig.~\ref{fig:2D_scaling}(c-f). We observe very different mode profiles; cross-stitch eigenmodes agree with our intuition, consisting of a collection of well-separated resonant CLS, weakly coupled via Anderson-localized dispersive states. In contrast, this weak coupling assumption is violated in the 2D pyrochlore; many CLS are excited in close proximity and they can no longer be considered independently. Similar behaviour also occurs in 2D Lieb and Kagome lattices which also host protected band crossings.

This strong coupling between resonant CLS is due to the overcompleteness of the flat band; the additional eigenstates are \emph{not} compactly localized, but instead wrap around the entire lattice, forming line states~\cite{bergman2008}. The coupling can be quantified by considering the range of the (real space) flat band projection operator, which determines how the CLS respond to a local perturbation. In regular flat bands the projection operator is either local ($U=1$) or short-ranged ($U>1$); the disorder perturbs each CLS independently. However in $d>1$ lattices with protected crossings, the projector is long-ranged with a power law decay $\sim 1/r^d$, leading to strong coupling, geometric frustration, and multifractality in the $W \rightarrow 0$ limit~\cite{chalker10}; the CLS respond to generic perturbations in a complex way. We stress however that under suitable correlated perturbations that for example do not shift the energy of some fraction of the CLS, weakly coupled CLS can persist~\cite{FB_correlations}.

\section{Conclusions}
\label{conclusion}

We have studied the localization of weakly disordered flat band eigenmodes in a variety of flat band lattices, showing that the flat band's compact localized states provide a useful way to understand its response to perturbations. When the flat band is separated from other dispersive Bloch bands by a gap the compact localized states are robust against disorder, retaining their strong localization under finite disorder. When the band gap vanishes, a Fano resonance-like effect enhances the effective disorder potential, leading to strong sensitivity to the disorder and different power law scalings $\nu$ of the Anderson localization length. We have verified the applicability of this picture to: (i) lattices with multiple flat bands, (ii) flat bands of class $U>1$, where the minimal compact localized states occupy multiple unit cells, and (iii) lattices with off-diagonal (hopping) disorder. 

In most cases, a reduction of the disordered eigenvalue problem to the exactly-solvable Lloyd model correctly predicts the scaling exponent $\nu$ of the localization length in 1D. While this reduction inexplicably fails for the 1D Lieb lattice flat band, for which $\nu \approx 4/3$, we found a reduction of this model to a previously studied case sharing the same anomalous exponent: the diamond ladder~\cite{fb0}. The appearance of $\nu = 4/3$ in dispersive bands in the Lieb lattice suggests a universal mechanism, probably related to the density of states, which would be interesting to explore further in future studies, either by applying the semi-analytical approach recently introduced in Ref.~\cite{ge_arxiv}, or generalizing the older perturbative expansions of Refs.~\cite{derrida84,titov03,goldhirsch94} to disorder with a divergent variance. We leave a rigorous analytical derivation of this anomalous scaling law as an open problem. 

Generalizing to 2D lattices, we found that the response to disorder becomes sensitive the ``geometric frustration'' of the flat band. In certain ``unfrustrated'' lattices disorder weakly couples compact localized states via dispersive states, similar to the 1D case. On the other hand, in ``frustrated'' lattices the coupling between CLS becomes long-ranged and the simple picture of independent Fano resonances breaks down under generic uncorrelated disorder.

Finally, while we have focused on the case of static disorder potentials, our detangling procedure may also be useful in characterizing the response of flat bands to other perturbations, such as nonlinearities~\cite{huber2010,johansson2015,vicencio2013} or balanced gain and loss~\cite{chern2015,pt_flat,molina2015}.

\section*{Acknowledgements}

This research was supported by the Australian Research Council, the Singapore National Research Foundation under grant No. NRFF2012-02, and the Institute for Basic Science through Project Code (IBS-R024-D1).

\section*{Author contribution statement}

All authors have contributed significantly to the article.

\appendix

\section{Localization length}

In 1D lattices with only a single transmission channel (such as the sawtooth, pyrochlore, and stub), we can apply the simple transfer number method to obtain the localization length. This relies on the fact that a single scalar equation describes the growth rate of the eigenmodes. As an example, we consider the two-band approximation of Eq.~\ref{eq:fano_eliminated}. Dividing by the dispersive wave amplitude $p_n$, defining the amplitude ratio $R_n \equiv p_{n+1}/p_n$ and rearranging, we obtain the map
\begin{equation}
R_n = V_n + \frac{(\epsilon_n^-)^2}{E - E_{FB} - \epsilon_n^+} - E - \frac{1}{R_{n-1}}.
\end{equation}
For Anderson localized modes, $p_n \sim e^{n / \xi}$, where $\xi$ is the localization length, such that $R_n \sim e^{1/\xi}$ and the localization length is obtained as~\cite{kramer93}
\begin{equation}
\xi^{-1} = \lim_{N\rightarrow \infty} \frac{1}{N} \sum_{n=1}^N \ln |R_n|.
\end{equation}
For sufficiently large $N$ the summand is self-averaging and this expression converges to $\xi^{-1}$.

In lattices with multiple ($M>1$) transmission channels (eg. 1D Lieb and 2D lattices), each channel $j$ has a different localization length $\xi_j$. If one eliminates variables to obtain a single equation for an amplitude ratio $R_n$, the solution will only be sensitive to the fastest growing channel, corresponding to the smallest localization length $\mathrm{min}(\xi_j)$, whereas the transport properties of the lattice are determined by the \emph{largest} localization length. Thus, the transfer number method no longer works, and the more general transfer matrix method is required~\cite{kramer93,transfer_matrix}. Typically one writes the eigenvalue problem as
\begin{equation}
\hat{t}_n \psi_{n+1} = (\hat{\epsilon}_n - E) \psi_n - \hat{t}_{n-1}^{\dagger} \psi_{n-1},
\end{equation}
where $\psi_n$ encodes the sublattice amplitudes, $\hat{t}$ describes intercell hopping, and $\hat{\epsilon}$ accounts for intracell hopping and onsite potentials. This can be rewritten as
\begin{eqnarray}
\left( \begin{array}{c} \psi_{n+1} \\ \psi_n \end{array} \right) &= \left( \begin{array}{cc} t_n^{-1} (\hat{\epsilon}_n - E) & -t_n^{-1} t_{n-1}^{\dagger} \\ \hat{1} & 0  \end{array} \right) \left( \begin{array}{c} \psi_{n} \\ \psi_{n-1} \end{array} \right), \nonumber \\ 
&= \hat{T} \left( \begin{array}{c} \psi_{n} \\ \psi_{n-1} \end{array} \right),
\end{eqnarray}
where $\hat{T}$ is the transfer matrix and $\hat{1}$ is the identity matrix. For sufficiently large $N$ the localization lengths $\xi_j$ are obtained from the $M$ largest eigenvalues $\lambda_j$ of $\hat{T}^N$ as $1/\xi_j = \frac{1}{N} \ln |\lambda_j|$. Note that a naive direct computation of the matrix $T^N$ via matrix multiplication will suffer from numerical overflow, hence one must periodically orthonormalize columns of the partial product $\hat{T}^n$~\cite{kramer93}.

To numerically obtain the localization length in Sec.~\ref{sec:examples} we used system sizes of $N \sim 10^6$ unit cells, which is sufficient to achieve convergence to $\xi$ and approximately 100 times larger than the largest localization length obtained.

\end{document}